\documentclass[journal=ancac3,manuscript=article,layout=traditional]{achemso}

\usepackage[version=3]{mhchem} 
\usepackage{natbib}
\usepackage{upgreek}
\usepackage{color}
\setkeys{acs}{abbreviations = false, biblabel = brackets, etalmode = truncate, maxauthors = 20, keywords = true}

\author{Chia-Ju Chen}
\affiliation{Department of Physics, National Tsing Hua University, 300044 Hsinchu, Taiwan}
\author{Yu-Tung Lin}
\affiliation{Department of Physics, National Tsing Hua University, 300044 Hsinchu, Taiwan}
\author{Chieh-Lin Lee}
\affiliation{Department of Physics, National Tsing Hua University, 300044 Hsinchu, Taiwan}
\author{Nitin Kumar}
\affiliation{Department of Physics, National Tsing Hua University, 300044 Hsinchu, Taiwan}
\author{Hung-Chin Lee}
\affiliation{Department of Physics, National Tsing Hua University, 300044 Hsinchu, Taiwan}
\author{Yen-Hui Lin}
\affiliation{Department of Physics, National Tsing Hua University, 300044 Hsinchu, Taiwan}
\author{Bo-Yao Wang}
\affiliation{Department of Physics, National Changhua University of Education, Changhua 500, Taiwan}
\author{Stefan Blügel}
\affiliation{Peter Grünberg Institut, Forschungszentrum Jülich and JARA, D-52425 Jülich, Germany}
\alsoaffiliation{Institute for Theoretical Physics, RWTH Aachen University, 52056 Aachen, Germany}
\author{Gustav Bihlmayer}
\email{g.bihlmayer@fz-juelich.de}
\affiliation{Peter Grünberg Institut, Forschungszentrum Jülich and JARA, D-52425 Jülich, Germany}
\author{Pin-Jui Hsu}
\email{pinjuihsu@phys.nthu.edu.tw}
\affiliation{Department of Physics, National Tsing Hua University, 300044 Hsinchu, Taiwan}
\alsoaffiliation{Center for Quantum Technology, National Tsing Hua University, Hsinchu 300044, Taiwan}

\title{Magnetic Triple-q State in Antiferromagnetic Monolayer Interfaced with Bismuthene}

\abbreviations{IR,NMR,UV}
\keywords{\textit{single-atomic-layer bismuthene, antiferromagnetic Mn monolayer, magnetic triple-q state, unidirectional magnetic anisotropy, spin-polarized scanning tunneling microscopy/spectroscopy, density functional theory}}

\begin{document}


\begin{abstract}

\textcolor{black}{We have successfully fabricated the bismuthene covered Mn monolayer on Ag(111) by evaporating Mn atoms onto ($p\times\sqrt{3}$)-Bi/Ag(111 at room temperature. By using spin-polarized scanning tunneling microscopy (SP-STM), we have resolved the magnetic triple-\textbf{q} (3Q) state. In combination with density-functional theory (DFT) calculations, the 3Q$^{3}$-like spin texture is the magnetic ground state for the bismuthene covered Mn monolayer/Ag(111). Interestingly, the uniaxial magnetic anisotropy of 3Q$^{3}$ state triggered by the bismuthene on top of Mn monolayer/Ag(111) has been revealed, which is consistent with the switching of 3Q$^{3}_{\text{up}}$ and 3Q$^{3}_{\text{down}}$ domains observed by SP-STM measurements with external magnetic fields.}
\end{abstract}
\section{Introduction}

\textcolor{black}{Noncollinear spin structures, such as chiral domain wall, spin spiral and magnetic skyrmion etc., are of significant importance in not only fundamental research studies, but also advanced technological applications\cite{EYVedmedenko,MBode,SSParkin,NRomming,NNagaosa,AFert,KvB,YYoshimura,PJHsu,ANBogdanov,YTakeuchi,GChen}. Under the framework of classical Heisenberg model, a magnetic spin spiral can be characterized by a single vector \textbf{q} in the reciprocal space of Brillouin zone and the superposition of three symmetry-equivalent single-\textbf{q} (1Q) spin spirals can result in a noncollinear magnetic order of triple-\textbf{q} (3Q) state\cite{PKurz,TOkubo,SHaldar,MGutzeit,FNickel}. In addition to the 2D solid $^{3}$He films on graphite\cite{TMomoi}, the magnetic 3Q state was also predicted in the monolayer (ML) Mn on Cu(111) surface more than two decades ago, constituting a three-dimensional (3D) noncollinear spin structure placed onto a 2D hexagonal lattice\cite{PKurz}. Until recently, the magnetic 3Q state has been experimentally observed on the hcp-stacked ML Mn/Re(0001) by spin-polarized scanning tunneling microscopy (SP-STM)\cite{JSpethmann} as well as on the triangular antiferromagnetic compounds Co$M_{3}$S$_{6}$ ($M$=Nb, Ta) by neutron scattering experiments\cite{HTakagi,PPark}. Intriguingly, a considerable topological orbital moment (TOM) can be induced from non-coplanar 3Q spin configuration that plays an essential role in developing spontaneous topological Hall effect without a need of spin-orbit interaction (SOI)\cite{MHoffmann,JPHanke,SGrytsiuk}. On top of that, the emergence of topological superconductivity has been proposed theoretically when a magnetic 3Q spin structure interfaced with the superconducting layers with \textit{s}-wave pairing symmetry\cite{JBedow}.}

\textcolor{black}{After the milestone discovery of single-atomic-layer graphene\cite{KSNovoselov,KSNovoselov1,AHCastroNeto}, extensive research interests have been stimulated to investigate the two-dimensional (2D) graphene-like materials that are classified as Xenes ranging from the group IV to VI atoms\cite{MEzawa,CLin,ZLLiu,EBianco,SBalendhran,LMatthes,FReis,AJMannix,QLiu,LXian,CGrazianetti,SMBeladi,GBihlmayer}. Among them, the last group-V element bismuth (Bi) can be synthesized into a heavy graphene-analogue of bismuthene that exhibits a wide range of exceptional physical and chemical properties. It is particularly noteworthy that bismuthene is made of Bi atoms with a high atomic mass (Z = 83), possessing a strong SOI for opening a topologically nontrivial gap $\sim$ 0.8 eV much higher than that of freestanding graphene\cite{FReis,RStuehler}. Through the substrate-orbital-filtering effect, the topological gap of bismuthene can be further increased up to about 1.0 eV, which is currently the largest value as compared to many other 2D monoelemental Xenes\cite{SSun}. Such a large topological gap turns bismuthene into one vital candidate of 2D topological insulators (TIs) for realizing quantum spin Hall effect (QSHE) at room temperature (RT)\cite{FReis,RStuehler}, leading to the possible ultrafast electron dynamics contributed by topologically protected edge states and the collective excitonic transitions occurred at RT, respectively\cite{JMaklar,MSyperek}.}

\textcolor{black}{In recent years, introducing magnetic ingredients to the 2D Xenes has received much attention in achieving novel functionalities in nanoelectronic and spintronic devices. For example, a wealth of proximity-driven spin-dependent properties have been uncovered in the graphene interfaced with ferromagnet and antiferromagnet heterostructures\cite{ZQiao,ZWang,GSong,BZhou,WHu}. Additionally, an intrinsic long-range magnetic order has been established in the silicene- and germanene-based stoichiometric layered compounds by the chemisorption of lanthanide ions\cite{AMTokmachev,DVAveryanov,AVMatetskiy}. Moreover, the spin pumping phenomenon and the coexistence of topological band structures and ferromagnetism have been illustrated in the low-temperature (LT) growth of stanene atomic layers on ferromagnetic Co films, providing one route toward observing quantum anomalous Hall effect (QAHE)\cite{CZChang,YTokura,BABernevig,LGladczuk1,CJChen}. In opposite to different Xenes mentioned above, merely theoretical predictions have been reported on the electronic and spin transport of bismuthene doped by magnetic impurities, but the relevant experimental attempts are still lacking. Therefore, it becomes an essential issue to experimentally explore the bismuthene coupled with magnetic materials, which would be indispensable to shed a light onto the mutual interplay between bismuthene and magnetism.}

\section{Results and discussion}

\begin{figure}[!ht]
  \includegraphics[width=8.5cm]{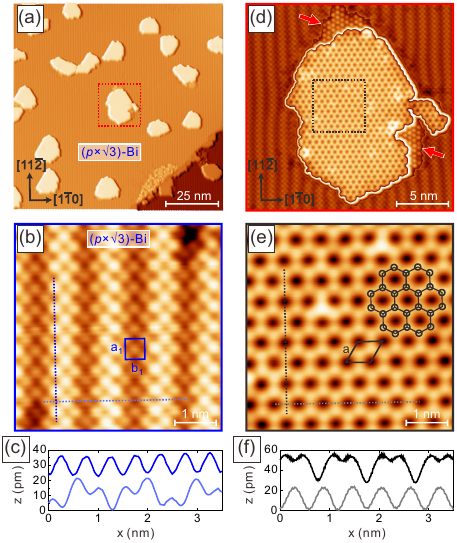}
  \caption{\textcolor{black}{(a)\,Overview of the STM constant-current topography of 0.15$\pm$0.05 ML Mn grown on ($p\times\sqrt{3}$)-Bi/Ag(111) at RT ($U_{b} = +0.1$\,V, $I_{t} = 0.4$\,nA). (b)\,Atomic resolution image of ($p\times\sqrt{3}$)-Bi/Ag(111), where the lattice constants of a$_{1}$ = 5.0$\pm$0.1 \AA\, and b$_{1}$ = 4.7$\pm$0.1 \AA\, have been extracted from the corresponding topographic line profiles plotted in (c) ($U_{b} = +10$\,mV, $I_{t} = 1.0$\,nA). (d)\,Zoom-in image acquired from the red dashed square of (a). The two red arrows indicate the ($\sqrt{3}\times\sqrt{3})R$30$^\circ$-BiAg$_{2}$ surface alloy nearby the boundary of Mn-induced superstructure ($U_{b} = +0.1$\,V, $I_{t} = 0.4$\,nA). (e)\,Atomically-resolved image on the Mn-induced superstructure from the black dashed square of (d). The individual atoms of honeycomb lattice are marked by black circles ($U_{b} = +10$\,mV, $I_{t} = 1.0$\,nA). (f)\,Topographic line profiles measured from black and gray dashed lines of (e), where the average bond length about 3.3$\pm$0.1 \AA\, (black line) between two adjacent atoms and the lattice constant of a = 5.8$\pm$0.1 \AA\, (gray line) have been obtained for the Mn-induced honeycomb structure.}}
  \label{fgr:Topo}
\end{figure}

\textcolor{black}{Fig. 1(a) represents the STM topographic overview of 0.15$\pm$0.05 ML Mn deposited onto ($p\times\sqrt{3}$)-Bi/Ag(111) at RT. Note that the surface coverage of Mn is calibrated from the atomic density of its epitaxial growth on Ag(111)\cite{CLGao}. It is also denoted that the dealloying phase, ($p\times\sqrt{3}$)-Bi, appears on Ag(111) when the Bi coverage is higher than the 0.33 ML of a ($\sqrt{3}\times\sqrt{3})R$30$^\circ$-BiAg$_{2}$ substitutional surface alloy, which can be explained by a compressive strain from incorporating large-size Bi atoms into Ag lattice and a lack of Bi and Ag bulk miscibility\cite{CKao,KHLZhang,JPHu}. Fig. 1(b) displays the atomic resolution image of ($p\times\sqrt{3}$)-Bi/Ag(111) in which the moir\'{e} pattern, i.e., a periodic stripe-like modulation along high symmetry [11$\bar{2}$] direction, arises from the interference between ($p\times\sqrt{3}$)-Bi overlayer and underneath Ag substrate\cite{CKao,KHLZhang,JPHu}. We would like to denote that three rotationally equivalent domains of ($p\times\sqrt{3}$)-Bi structure can coexist due to a three-fold rotational symmetry on the Ag(111) surface \textcolor{black}{(see Supplementary Fig. S1 for details)} and the centered rectangular unit cell of ($p\times\sqrt{3}$)-Bi structure has been marked by the blue rectangular frame in Fig. 1(b). Fig. 1(c) summarizes the topographic line profiles measured from Fig. 1(b) along the [11$\bar{2}$] (blue dashed line) and [1$\bar{1}$0] (light blue dashed line) crystalline axes, respectively. The corresponding atomic lattice constants of a$_{1}$ = 5.0$\pm$0.1 \AA\, and b$_{1}$ = 4.7$\pm$0.1 \AA\, have been extracted for the ($p\times\sqrt{3}$)-Bi/Ag(111), which are in line with previous studies\cite{CKao,KHLZhang,JPHu}.}

\textcolor{black}{Fig. 1(d) is the zoom-in topographic image acquired from Fig. 1(a) (red dashed square), where the 2D ordered superstructure has been formed on top of ($p\times\sqrt{3}$)-Bi/Ag(111) after Mn deposition at RT. It is worth mentioning that the growth of a Mn-induced superstructure has been hindered when evaporating Mn onto ($p\times\sqrt{3}$)-Bi/Ag(111) at 80 K \textcolor{black}{(see Supplementary Fig. S2 for details)}. In addition, there are small patches of ($\sqrt{3}\times\sqrt{3})R$30$^\circ$-BiAg$_{2}$ surface alloy in close proximity to the Mn-induced superstructure as marked by two red arrows in Fig. 1(d) \textcolor{black}{(see Supplementary Fig. S3 for details)}, indicative of a local rearrangement of Bi atoms. Interestingly, the 2D honeycomb (HC) lattice of the Mn-induced superstructure has been atomically resolved in Fig. 1(e) that is directly measured from the black dashed square in Fig. 1(d). Since the individual atoms have the same size and shape in Fig. 1(e), one would expect such uniformity originated from a single constituent, i.e., either Bi or Mn atoms, in the Mn-induced HC superstructure. From the topographic line profiles plotted in Fig. 1(f), i.e., gray and black dashed lines from Fig. 1(e), the lattice constant of a = 5.8$\pm$0.1 \AA\, (gray line) has been extracted for the Mn-induced HC structure, resulting in the  $(2 \times2)$ supercell with respect to the  $(1 \times 1)$ unit cell of Ag(111). Furthermore, the atomic flatness of the Mn-induced HC structure can be referred to the atomic corrugation height on nonequivalent sites from the topographic line profile (black line) in Fig. 1(f), and the average bond length about 3.3$\pm$0.1 \AA\,  between two adjacent atoms has been obtained, which is also in agreement with the lattice periodicity of a $(2 \times2)$ supercell.}

\begin{figure}[!ht]
  \includegraphics[width=8.5cm]{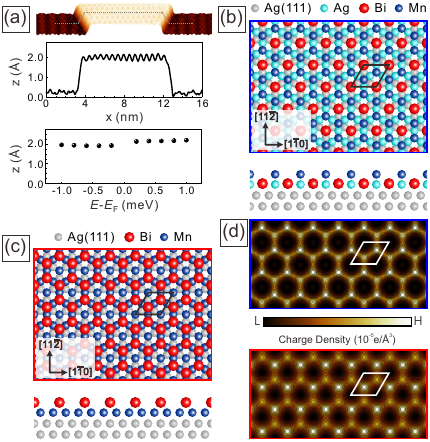}
  \caption{(a)\,\textcolor{black}{Top: STM topography of Mn-induced honeycomb structure on ($p\times\sqrt{3}$)-Bi/Ag(111) ($U_{b} = +10$\,mV, $I_{t} = 1.0$\,nA). Middle: topographic line profile taken from the gray dashed line in top panel. A monolayer height of 1.9$\pm$0.1 \AA\, has been obtained. Bottom: bias-dependent heights in the broad range of ±1.0 V. (b)\,Structure model of a Mn honeycomb lattice on ($2 \times 2$)-BiAg$_{3}$/Ag(111) for top and side views. (c)\,Structure model of bismuthene on p$(1 \times 1)$-Mn/Ag(111) for top and side views. Black rhombus marks the $(2 \times2)$ supercell in the inset. (d)\,Simulated STM images of Mn honeycomb lattice on $(2 \times2)$-BiAg$_{3}$/Ag(111) (top) and bismuthene on $(1 \times 1)$-Mn/Ag(111) (bottom). White rhombus marks the $(2 \times 2)$ supercell in the inset. The Mn honeycomb lattice and bismuthene turn out having a trivial deviation in topography, which is difficult to be explicitly identified in terms of STM measurements alone.}}
  \label{fgr:STS}
\end{figure}

\textcolor{black}{According to the topographic line profile taken across the Mn-induced HC structure in the top panel of Fig. 2(a), it has an apparent height about 1.9$\pm$0.1 \AA\, on ($p\times\sqrt{3}$)-Bi/Ag(111) at the sample bias voltage of +10 mV, suggesting a film thickness of single atomic layer. Besides the measurement at +10 mV, the apparent height has also been examined in the wide range of bias voltages. As summarized in the bottom panel of Fig. 2(a), the average height is 2.1$\pm$0.1 \AA\, within the bias range of $\pm$1.0 V \textcolor{black}{(see Supplementary Fig. S4 for details)}, which is slightly lower but comparable to a single atomic step height 2.36 \AA\, of Ag(111) from the simple hard-sphere model\cite{MJHarrison,MFCrommie1}. Based on the monoatomic step height in Fig. 2(a) as well as the elemental homogeneity in Fig. 1(e), several structure models, including a full overlayer, a mixture of Mn and Bi alloy, different atomic stackings and potential lattice transformations etc., have been considered for the Mn-induced HC structure. Among all of them, we have found two structure models that are able to be fully relaxed and stabilized from the DFT calculations. As illustrated in the top panel of Fig. 2(b), the Mn honeycomb lattice can be constructed on the surface alloy BiAg$_{3}$/Ag(111), where the (2x2) supercell can be inferred from the black rhombus in the inset. \textcolor{black}{Incidentally, the (antiferromagnetic) Mn HC lattice atop of p($2\times2$)-BiAg$_{3}$/Ag(111) also fulfills the prerequisite of monoatomic step height, \textcolor{black}{i.e., about 2.24 \AA\ extracted from the DFT structure relaxations,} as shown in the side view of bottom panel of Fig. 2(b).}}

\textcolor{black}{Apart from the Mn HC lattice/($\sqrt{3}\times\sqrt{3})R$30$^\circ$-BiAg$_{2}$/Ag(111), the other energetically favorable structure of a Bi honeycomb lattice, i.e., bismuthene, covered on p$(1 \times 1)$-Mn/Ag(111) has been presented in Fig. 2(c). In this case, the deposited Mn atoms diffuse underneath and develop into a close-packed Mn ML with  $(1 \times 1)$ unit cell on the Ag(111) surface. Note that a Mn ML following the fcc stacking as a result of pseudomorphic growth has been reported on Ag(111)\cite{CLGao}. On the other hand, the Bi atoms have been pushed upward and reassemble into single-atomic-layer bismuthene, where the $(2 \times2)$ supercell has been designated by the black rhombus in the top panel of Fig. 2(c). As can be seen from the side view of bottom panel of Fig. 2(c), this structure model requires the diffusion process of Mn atoms, elucidating why it is more likely to happen at the growth temperature of RT rather than of 80 K \textcolor{black}{(see Supplementary Fig. S2 for details)}. In addition, for the two stable structures calculated from DFT, the} \textcolor{black}{Mn honeycomb structure on BiAg$_3$/Ag(111) and the Bi honeycomb lattice on Mn/Ag(111), we calculated STM images taking all states in a bias window from the Fermi-level, $E_{\mathrm F}$, to $E_{\mathrm F}+0.5 eV$. As can be seen in Fig.~2(d), in both simulations a bright HC lattice appears with some contrast between the two sublattices. 
}

\begin{figure}[!ht]
  \includegraphics[width=\textwidth]{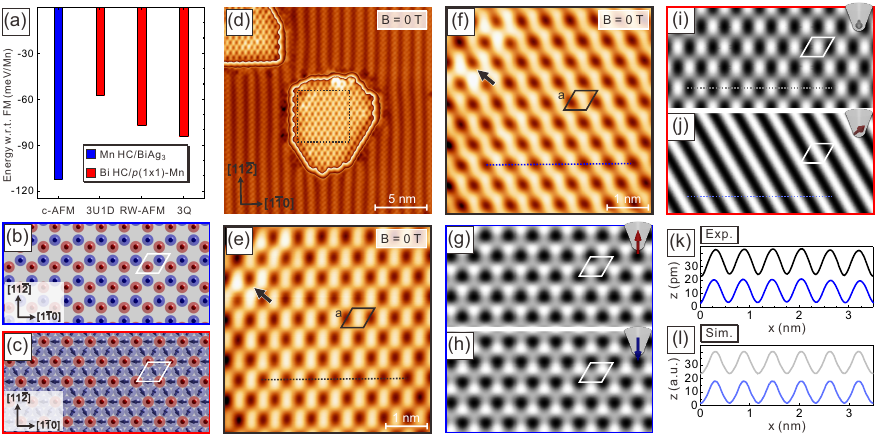}
  \caption{(a)\,\textcolor{black}{Energy comparison of different magnetic states with respect to ferromagnetism. (b)\, and (c)\,Atomic spin structures of c-AFM order and magnetic 3Q state found in the Mn honeycomb lattice on $(2 \times 2)$-BiAg$_{3}$/Ag(111) and the bismuthene on $(2 \times 2)$-Mn/Ag(111). (d)\,SP-STM topography of as-grown sample measured by bulk Cr tip in zero magnetic field. (e)\,Atomic-scale magnetic image acquired from the black dashed square of (d), where the checkerboard spin contrast forms a $(2 \times2)$  magnetic superlattice (black rhombus). (f)\,Magnetic stripe pattern obtained by the second Cr tip magnetization. Black arrow refers to a landmark at the same scan area with (e). (g)\,, (h)\,Simulated spin contrasts of the c-AFM order from (b) by employing the out-of-plane tip magnetizations parallel and antiparallel to the N\'{e}el vector. (i)\,, (j)\,Simulated checkerboard and stripe spin contrasts of magnetic 3Q state from (c) with the tip magnetizations depicted discretely in the insets. White rhombus marks the $(2 \times2)$  supercell in the inset. (k)\, (l)\, Comparison of color-coded line profiles measured from experiments ((e) and (f)) and simulations ((i) and (j)), respectively. The magnetic period of 5.8$\pm$0.1 \AA\, has been extracted along the high symmetry [1$\bar{1}$0] direction, which is in line with theoretical simulations. (scan parameters: $U_{b} = +10$\,mV, $I_{t} = 1.0$\,nA)}}
  \label{fgr:QPI}
\end{figure}

\textcolor{black}{The two stable structures, discovered from DFT calculations, appear nearly identical and can not be clearly distinguished on the basis of STM measurements and simulations. However, their magnetic ground states have been further investigated and compared energetically with respect to the ferromagnetism.} \textcolor{black}{In the Mn HC lattice, we find a clear preference for an antiferromagnetic (AFM) order, $112$~meV per Mn atom  lower in energy than the ferromagnetic (FM) state. In the Bi HC structure, we have different possibilities of AFM ordering in the p$(2 \times2)$ Mn unit cell, e.g.\ collinear ones like a row-wise AFM order, noncollinear ones like the 3Q state that was predicted as ground state for Mn/Cu(111)~\cite{PKurz} or also ferrimagnetic ones with three spins up and one down (3u1d). As can be seen from Fig.3(a), the 3Q state has the lowest energy of these, with $-84$~meV/Mn energy difference to the FM state. 
} 
\textcolor{black}{Fig. 3(b) and (c) show the atomic spin textures of c-AFM order and the magnetic 3Q state with the out-of-plane magnetization component color-coded, which can be identified by utilizing SP-STM technique with a high spatial and spin resolution down to atomic scale.\cite{KvB,RW,MB}........................} 

\textcolor{black}{Fig. 3(d) represents the SP-STM topographic overview of as-grown Mn-induced honeycomb structure on ($p\times\sqrt{3}$)-Bi/Ag(111) measured by a bulk Cr tip in the absence of an external magnetic field. Fig. 3(e) is the atomically-resolved magnetic image on the Mn-induced honeycomb structure acquired from the black dashed square of Fig. 3(d), where the checkerboard pattern has been revealed instead of honeycomb lattice observed by a nonmagnetic tip in the Fig.~1(e), insinuating its magnetic origin. On top of that, the checkerboard spin texture forms a (2x2) magnetic superlattice, i.e., it has the same periodicity as the (2x2) supercell of the honeycomb lattice, given by the black rhombus inferred from the inset of Fig.~3(e). Since the antiferromagnetic Cr material is effectively immune to an external magnetic field, a small voltage pulse has been applied to the bulk Cr tip to modify the magnetization direction at the tip apex. Note that such tip voltage pulse method has been conducted at the sample area distant away to avoid the local surface contamination. By means of this second Cr tip magnetization direction, the magnetic stripe-like feature has been unveiled in Fig.~3(f), which constitutes the other characteristic magnetic pattern aside from magnetic checkerboard structure in Fig.~3(e). Note that the scan position of Fig.~3(f) is identical to the one in Fig.~3(e) in which a landmark presumably due to atomic electronic inhomogeneity can be referred to (black arrow). It is also denoted that the rotationally symmetric checkerboard and stripe patterns have been resolved on the same scan position by changing different Cr tip magnetizations \textcolor{black}{(see Supplementary Fig. S5 for details).}}

\textcolor{black}{To gain more insights on the SP-STM results by using bulk Cr tip,  the simulated SP-STM image of the c-AFM order from Fig.~3(b) has been demonstrated in Fig.~3(g) with an out-of-plane tip magnetization as depicted in the inset. Note that the out-of-plane tip magnetization is chosen to be parallel to the (N\'{e}el vector) of the c-AFM order, giving rise to the maximum output of intensity in the simulated SP-STM image of Fig.~3(g). On the contrary, the tip magnetization direction employed in Fig.~3(h) is opposite to that of Fig.~3(g), leading to a phase shift in the atomic spin contrast within honeycomb sublattice. Besides the c-AFM order, the SP-STM simulations have been further carried out on the magnetic 3Q state of Fig.~3(c). Interestingly, the checkerboard as well as the stripe patterns have been reproduced in Fig.~3(i) and (j), where magnetic tip direction and (2x2) magnetic superlattice (white rhombus) are designated separately in the inset. We would like to denote that simulated SP-STM images on both c-AFM and magnetic 3Q spin structures with systematically varying magnetic tip directions have been cross-checked \textcolor{black}{(see Supplementary Fig. S6 and Fig. S7 for details).} Moreover, the topographic line profiles taken from black and blue dashed lines in Fig.~3(e) and (f) have been plotted in Fig.~3(k), corresponding to a magnetic period of 5.8$\pm$0.1 \AA\, along the high symmetry [1$\bar{1}$0] direction. This is consistent with the value of 5.78 \AA\, from the simulated profiles in Fig.~3(l), which are taken from gray and light blue dashed lines in the Fig.~3(i) and (j), respectively. As the magnetic patterns, i.e., checkerboard and stripe, and magnetic periodicity can be restored from theoretical SP-STM simulations, the bismuthene covered \textit{p}(1x1)-Mn/Ag(111) not only validates the energetically preferred structure, but also supports the magnetic 3Q state for the experimental SP-STM observations with the bulk Cr tip.}

\begin{figure*}[!ht]
	\includegraphics[width=8.5cm]{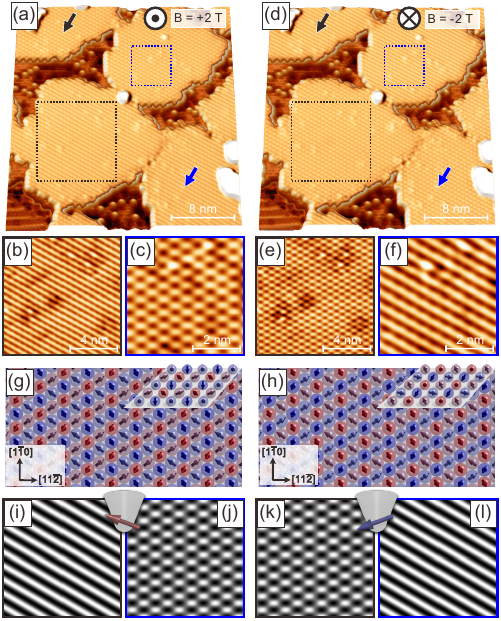}
	\caption{\textcolor{black}{(a)\,SP-STM topographic overview of magnetic checkerboard and stripe domains measured by Fe coated W tip at B$_{z}$ = +2 T. (b)\,,(c)\,Zoom-in images of magnetic checkerboard and stripe spin contrasts from black and blue dashed squares of (a). (d)\,Exchange of magnetic checkerboard and stripe domains revealed at B$_{z}$ = -2 T. Equivalent behavior also occurs at the other domains (black and blue arrows). (e)\,,(f)\,Atomically-resolved checkerboard and stripe spin contrasts from black and blue dashed squares of (d). (g)\,,(h)\, Atomic spin structures of magnetic 3Q$_{\text{up}}$ and 3Q$_{\text{down}}$ states. The perspective views are shown in the inset. (i)\,,(j)\,Simulated checkerboard and stripe spin contrasts from magnetic 3Q$_{\text{up}}$ and 3Q$_{\text{down}}$ states, respectively. Inset depicts the canted tip magnetization employed in the SP-STM simulations. (i)\,,(j)\,Magnetic checkerboard and stripe spin contrasts swap with each other when only the out-of-plane component of tip magnetization is inverted. (scan parameters: $U_{b} = +10$\,mV, $I_{t} = 1.0$\,nA).}
	}
	\label{fgr:DFT}
\end{figure*}

\textcolor{black}{After understanding the zero-field measurements, the field-dependent SP-STM experiments have been performed using a ferromagnetic Fe-coated W tip in order to switch the tip magnetization direction accordingly\cite{KvB,RW,MB}. Fig.~4(a) shows the real-space spin map on the bismuthene covered \textit{p}(1x1)-Mn/Ag(111) with the tip magnetization that can be aligned along the out-of-plane sample direction at B$_{z}$ = +2 T. Fig.~4(b) and (c) are the zoom-in images acquired from black and blue dashed squares in the Fig.~4(a), where the characteristic checkerboard and stripe spin contrasts have been resolved. Note that we did not observe rotational domains, suggesting the atomic spin structures of magnetic checkerboard and stripe patterns are symmetrically mirrored along out-of-plane direction and manifest themselves as magnetic 3Q$_{\text{up}}$ and 3Q$_{\text{down}}$ domains. \textcolor{black}{(see Supplementary Fig. S8 for details).} When applying an external magnetic field B$_{z}$ of $-2$ T, the resultant magnetic image with an opposite out-of-plane tip magnetization has been displayed in Fig.~4(d). Intriguingly, the magnetic checkerboard and stripe patterns exchange with each other as the magnified images shown in Fig.~4(e) and (f), i.e., black and blue dashed squares from the Fig.~4(b), implying the presence of uniaxial magnetic anisotropy in the bismuthene covered \textit{p}(1x1)-Mn/Ag(111) system. The equivalent behavior also occurs at the other magnetic domains, for example, as indicated by black and blue arrows by comparing the Fig.~4(a) with the Fig.~4(d). Note that the external magnetic field has been swept consecutively from B$_{z}$ = +2 T to -2 T to verify the switching between magnetic checkerboard and stripe domains \textcolor{black}{(see Supplementary Fig. S9 for details)}.}

\textcolor{black}{As reported from previous literature\cite{JSpethmann,SHaldar,FNickel}, three highly symmetric 3Q spin configurations, i.e., 3Q$^{\text{i}}$, \text{i} = 1, 2, 3, can exist and transform into one another through a rotation of all spins in the 2D magnetic ultrathin film. Note that the tetrahedral angle between all adjacent spins of an ideal 3Q state is $\uptheta_{\text{t}}$ = cos$^{-1}(-\frac{1}{3})$ $\approx$ 109.5$^{\circ}$. By rotating all spins with $\frac{\uptheta_{\text{t}}}{2}$ and then $\frac{\pi}{2}$, the continuous transformation from 3Q$^{1}$ to 3Q$^{2}$ and further from 3Q$^{2}$ to 3Q$^{3}$ can be achieved~\cite{JSpethmann}. To explain the magnetic domain swapping from field-dependent SP-STM measurements, the magnetic 3Q$^{1}$, 3Q$^{2}$ and 3Q$^{3}$ spin structures have all been evaluated in the SP-STM simulations under the constraint of out-of-plane tip magnetization inversion \textcolor{black}{(see Supplementary Fig. S10 to Fig. S12 for details).} After the detailed analyses on the stimulated SP-STM results, the 3Q$^{3}$-like spin texture (3Q$^{3}_{\text{up}}$) in Fig.~4(g) and its counterpart with a reversed out-of-plane spin component (3Q$^{3}_{\text{down}}$) in Fig.~4(h) have been deduced, reconciling with the experimental observations from both Fig.~3 and Fig.~4. As demonstrated in Fig.~4(i) and (j), magnetic checkerboard and stripe patterns can be successfully generated from 3Q$^{3}_{\text{up}}$ (Fig.~4(g)) and 3Q$^{3}_{\text{down}}$ (Fig.~4 (h)) with a canted tip magnetization direction, i.e., consisting of both in-plane and out-of-plane magnetization components, as depicted in the inset. Most importantly, the exchange of magnetic checkerboard and stripe patterns has been fulfilled in Fig.~4(k) and (l) upon merely inverting the out-of-plane tip magnetization, yielding a good agreement with Fig.~4(a) and (d).}

\begin{figure*}[!ht]
	\includegraphics[width=8.5cm]{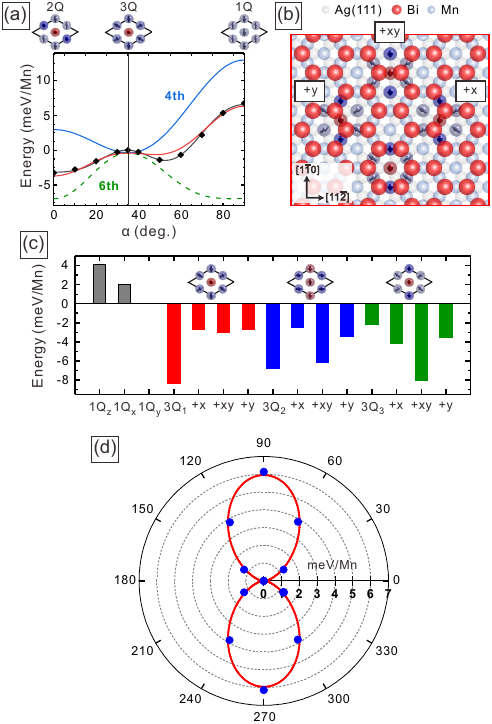}
	\caption{(a) \textcolor{black}{Total energy (without SOC) along a continuous transformation path between the 2Q, 3Q and 1Q state of a bismuthene covered Mn/Ag(111) surface: black diamonds: DFT results, black curve: fit to contributions of higher-order interactions up to $10^{\mathrm{th}}$ order. Contributions to $4^{\mathrm{th}}$ and $6^{\mathrm{th}}$ order and their sum are indicated by blue. green and red lines, respectively. The energies are referenced to the 3Q state. (b) Different positions of the bismuthene w.r.t.\ the 3Q$^3$ state of Mn/Ag(111). $+x$, $+y$ and $+xy$ indicate shifts of the lower configuration in different lattice vector directions. The total energy (including SOC) of three different 3Q states with bismuthene in these positions is shown in (c) with respect to the energy of the 1Q state with magnetization in $y$-direction (1Q$_y$). The energies of other 1Q states with magnetization in $x$ and $z$ direction are shown by the black bars. Panel (d) shows the total energy for the 3Q$^3+xy$ state when the spins are rotated around the $z$-axis. Due to the position of the bismuthene, an uniaxial anisotropy appears.}
	}
	\label{fgr:DFT}
\end{figure*}

\textcolor{black}{We performed additional DFT calculations to rationalize the experimental findings.  In a first step, we calculated the total energy of single-, double-, and triple-Q structures without spin-orbit coupling (SOC) contributions. In Fig.~5(a) we plot them along a continuous path connecting these states. Since Heisenberg-type (two-spin) contributions stay constant on this path, the energy variation can be decomposed in four-spin interactions and higher order terms~\cite{SHaldar}. Like in Mn/Re(0001), the fourth-order term stabilizes the 3Q state, but the sixth-order interactions make it a local maximum again. Both, a distorted 3Q state and the 2Q state form minima that are $2-3$~meV below the 3Q state. For the analysis of the contribution from SOC, we start again from the 3Q state. As mentioned above, different orientations of the 3Q state w.r.t.\ the lattice lead to different energies. Like in Ref.~\cite{JSpethmann}, we investigate the highly symmetric 3Q$^{i}$ states ($i=1,2,3$) but now they can have different positions relative to the Bi honeycomb lattice (Fig.~5(b)). Shifting the magnetic lattice half a lattice vector of the p$(2 \times 2)$ unit cell in $x$, $y$, or both directions results in structures denoted by 3Q$^{i}+x$, 3Q$^{i}+y$3, and Q$^{i}+xy$ respectively. We see from panel (c) that these shifts result in large energy variations up to 5~meV. The total energies are here referenced to the collinear 1Q state, with spins pointing in the easy $[11\overline2]$-direction (1Q$_y$). The $z$ and $[1\overline10]$ axes are the hard and medium directions for the 1Q state. We note, however, that the in-plane anisotropy of the 3Q states can be uniaxial. As shown in Fig.~5(d), uniform rotation of the spins of the 3Q$^3+xy$ state show a clear preference for the orientation of the spins along a single axis. This can explain the absence of orientational domains in the experiments.}

\section{Conclusions}

\textcolor{black}{We have successfully fabricated the bismuthene covered Mn monolayer on Ag(111) by evaporating Mn atoms onto $p\times\sqrt{3}$)-Bi/Ag(111 at room temperature. By using spin-polarized scanning tunneling microscopy (SP-STM), we have resolved the magnetic triple-\textbf{q} (3Q) state. In combination with density-functional theory (DFT) calculations, the 3Q$^{3}$-like spin texture is the magnetic ground state for the bismuthene covered Mn monolayer/Ag(111). Interestingly, the uniaxial magnetic anisotropy of 3Q$^{3}$ state triggered by the bismuthene on top of Mn monolayer/Ag(111) has been revealed, which is consistent with the switching of 3Q$^{3}_{\text{up}}$ and 3Q$^{3}_{\text{down}}$ domains observed by SP-STM measurements with external magnetic fields.}

\section{Methods}

\subsection{Sample preparation}

The Mn on ($p\times\sqrt{3}$)-Bi/Ag(111) was prepared in an ultrahigh vacuum (UHV) chamber with the base pressure below $p \leq 2 \times 10^{-10}$\,mbar. The Ag(111) surface was first prepared by cycles of Ar$^{+}$ ion sputtering with an ion energy of 500\,eV at room temperature and subsequent annealing up to 900\,K. The Bi source with purity of 99.999\,\% (Goodfellow) was sublimated from a molybdenum (Mo) crucible in an e-beam evaporator (FOCUS) while keeping the clean Ag(111) substrate at 450\,K. After that, high purity Mn (99.995\%, Goodfellow) was e-beam evaporated onto the ($p\times\sqrt{3}$)-Bi/Ag(111) at RT. As results of atomic interdiffusion and structural rearrangement, the bismuthene covered on \textit{p}(1x1)-Mn/Ag(111) with a $(2\times2)$ lattice periodicity has been fabricated.

\subsection{SP-STM/SP-STS measurements}

The SP-STM measurements were performed by using either chemically etched bulk Cr tip or Fe coated W tip in the LT-STM (UNISOKU, USM-1500) setup. For topographic images, STM was operated in the constant-current mode with the bias voltage $U$ applied to the sample. For tunneling spectroscopy (STS) measurements, a small bias voltage modulation was added to $U$ (frequency $\nu = 3991$\,Hz), such that tunneling differential conductance $\mathrm{d}I/\mathrm{d}U$ spectra and mappings can be acquired by detecting the first harmonic signal from a lock-in amplifier. External magnetic field up to 8 T was available to be applied to the sample along the surface normal direction.

\subsection{DFT calculations}

\textcolor{black}{
We performed DFT calculations using the full-potential linearized augmented plane-wave method as implemented in the {\sc Fleur} code~\cite{Fleur}. The symmetric film consisted of five Ag(111) layers with Mn and Bi overlayers embedded in semi-infinite vacuum~\cite{Krakauer:79}. For the relaxations, the generalized gradient approximation in the form of Perdew et al.~\cite{PBE} was used, while the non-collinear calculations were performed in the GGA-optimized structure using the local density approximation~\cite{Vosko:80}. In the non-collinear calculations~\cite{Kurz:04}, asymmetric films with six Ag substrate layers were used and spin-orbit coupling was included self-consistently.
}


\begin{acknowledgement}

\textcolor{black}{G.B. gratefully acknowledges computing time granted through JARA-HPC on the supercomputer JURECA at Forschungszentrum Jülich.} P.J.H. acknowledges support from National Science and Technology Council (NSTC) of Taiwan under Grant Nos. NSTC-112-2636-M-007-006 and NSTC-112-2112-M-007-037, Ministry of Science and Technology (MOST) of Taiwan under Grant Nos. MOST-111-2636-M-007-007 and MOST-110-2636-M-007-006, and center for quantum technology from the featured areas research center program within the framework of higher education sprout project by Ministry of Education (MOE) in Taiwan.

\end{acknowledgement}

\begin{suppinfo}

\end{suppinfo}

\bibliography{Mn_Bi_Ag111_1stSubm_20250718}

\end{document}